%Paper: hep-th/9401098
%From: swapna@imsc.ernet.in (Swapna Mahapatra)
%Date: Thu, 20 Jan 94 14:56:47+050

\input phyzzx
\baselineskip 24pt plus 1pt minus 1pt
%%%%%%%%%%%%%%%%%%%%%%%%%%%%%%%%%%%%%%%%%%%%%%%%%%%
\def\p{\partial}
\def\bp{\bar\partial}
\def\d{\delta}
\def\t{\tilde}
%%%%%%%%%%%%%%%%%%%%%%%%%%%%%%%%%%%%%%%%%%%%%%%%%%%%
%%%%%%%%%%%%%%%%%%%%%%%%%%%%%%%%%%%%%%%%%%%%%%%%%%%%%%%%%%%%%%%%%%%%%%
\NPrefs
\def\define#1#2\par{\def#1{\Ref#1{#2}\edef#1{\noexpand\refmark{#1}}}}
\def\con#1#2\noc{\let\?=\Ref\let\<=\refmark\let\Ref=\REFS
         \let\refmark=\undefined#1\let\Ref=\REFSCON#2
         \let\Ref=\?\let\refmark=\<\refsend}

%%%%%%%%%%%%%%%%%%%%%%%%%%%%%%%%%%%%%%%%%%%%%%%%%%%%%%%%%%%%%%%%%%%%
%%%%%%%%%%%%%%%%%%%%    REFERENCES   %%%%%%%%%%%%%%%%%%%%%%%%%%%%%%%
%%%%%%%%%%%%%%%%%%%%%%%%%%%%%%%%%%%%%%%%%%%%%%%%%%%%%%%%%%%%%%%%%%%%

\define\GIBB
G. Gibbons and K. Maeda, Nucl. Phys. {\bf B298} (1988) 741;
D. Garfinkle, G. Horowitz and A. Strominger, Phys. Rev. {\bf D43}
(1991) 3140;
M. Mueller, Nucl. Phys. {\bf B337} (1990) 37;
G. Mandal, A. M. Sengupta and S. Wadia, Mod. Phys. Lett. {\bf A6}
(1991) 1685.

%G. Horowitz and A. Strominger, Nucl. Phys. {\bf B360} (1991) 197;
%A. Shapere, S. Trivedi and F. Wilczek, Mod. Phys. Lett.
%{\bf A6} (1991) 2677.

%\define\GIDD
%S. Giddings and A. Strominger, Phys. Rev. Lett.
%{\bf 67} (1991) 2930.

%\define\MYERS
%R. C. Myers, Nucl. Phys. {\bf B289} (1987) 701;
%R. C. Myers and M. Perry, Ann. Phys. {\bf 172} (1986) 304;
%C. Callan, R. C. Myers and M. Perry, Nucl. Phys. {\bf B311}
%(1988) 673.

%\define\DABH
%A. Dabholkar, G. Gibbons, J. Harvey anf F. R. Ruiz,
%Nucl. Phys. {\bf B340} (1990) 33.

%\define\CALL
%C. Callan, J. Harvey and A. Strominger, Nucl. Phys. {\bf B359}
%(1991) 611.

%\define\DUFF
%M. Duff and J. Lu, Phys. Rev. Lett. {\bf 66} (1991) 1402;
%Nucl. Phys. {\bf B354} (1991) 141.

\define\WITT
E. Witten Phys. Rev. {\bf D44} (1991) 314;

%\define\WIT
%E. Witten, Comm. Math. Phys. {\bf 92}(1984)455.

\define\DVV
R. Dijkgraaf, H. Verlinde and E. Verlinde, Nucl. Phys.
{\bf B 271}(1992)269.

%\define\SEN
%A. Sen, TIFR/TH/92-57 (hep-th/9210050);
%G. Horowitz, UCSBTH-92-32 (hep-th/9210119);
%J. Harvey and A. Strominger, EFI-92-41 (hep-th/9209055).

\define\HORO
J. Horne and G. Horowitz, Nucl. Phys. {\bf B368}(1992) 444.

%\define\ASEN
%A. Sen, Phys. Rev. Lett. {\bf 69} (1992), 1006.

\define\VENE
G. Veneziano, Phys. Lett. {\bf B265} (1991) 287;
K. Meissner and G. Veneziano, Phys. Lett. {\bf B267} (1991) 33;
M. Gasperini, J. Maharana and G. Veneziano, Phys. Lett. {\bf B272}
(1991) 277.

\define\ASHOKE
A. Sen, Phys. Lett. {\bf B271} (1991) 295;
{\it{ibid}} {\bf B274} (1991) 34;
S. F. Hassan and A. Sen, Nucl. Phys. {\bf B375} (1992) 103.

%\define\MICRO
%A. Sen, TIFR/TH/92-29, hep-th/9206016.

\define\KKK
S. Khastgir and A. Kumar, Mod. Phys. Lett. {\bf A6} (1991) 3365;
S. Khastgir and J. Maharana, Class. Quant. Grav. (1993).

\define\KUMAR
A. Kumar, Phys. Lett. {\bf B293} (1992) 49.

\define\GIVEON
A. Giveon and M. Rocek, Nucl. Phys. {\bf B380} (1992) 128.

\define\ROCEK
M. Rocek and E. Verlinde, Nucl. Phys. {\bf B373} (1992) 630.

\define\MAHARANA
J. Maharana and J. Schwarz, CALT-68-1790.

\define\HHS
J. Horne, G. Horowitz and A. Steif, Phys. Rev. Lett. {\bf 68}
(1992) 568.

\define\BUSH
T. Busher, Phys. Lett. {\bf B201} (1988) 466;
{\it{ibid}} {\bf B194} (1987) 59.

\define\ARDALAN
M. Alimohammadi, F. Ardalan and H. Arfaei, BONN-HE-93-12; SUTDP-93/72/3;
IPM-93-007 (hepth/9304024).

\define\KIRITSIS
E. Kiritsis, Mod. Phys. Lett. {\bf A6} (1991) 2871.

%\define\PANV
%J. Panvel, Phys. Lett. {\bf B284}(1992)50; A. Tseytlin, Mod. Phys. Lett.
%{\bf A6} (1991) 1721.

\define\TSEYT
A. Tseytlin, Nucl. Phys. {\bf B399}(1993) 601.

\define\KIRIT
C. Nappi and E. Witten, IASSNS-HEP-93/61 (hepth/9310112);
K. Sfetsos, THU-93-30 (hepth/9311010), THU-93-31 (hepth/9311093);
E. Kiritsis and C. Kounnas, CERN-TH.7059/93 (hepth/9310202);
D. I. Olive, E. Rabinovici and A. Schwimmer, SWA/93-94/15,
WIS-93/1-CS, RI-93/69.

\define\SWAP
S. Kar and A. Kumar, Phys. Lett. {\bf B 291}(1992)246;
S. Mahapatra, Mod. Phys. Lett. {\bf A7}(1992)2999;
S. Kar, S. P. Khastgir and G. Sengupta, Phys. Rev. {\bf D47}(1993)3643.

\define\BARS
I. Bars and K. Sfetsos, Phys. Rev. {\bf D48} (1993) 844;
A. Tseytlin, CERN preprint, hepth/9301015;
K. Sfetsos, USC-93/HEP-S1 (hepth/9305074);
K. Sfetsos and A. Tseytlin, CERN-TH.6962/93, (hepth/9308018).

\define\GAUME
X. de la Ossa and F. Quevedo, Nucl. Phys. {\bf B403}
(1993) 377;
A. Giveon and M. Rocek, ITP-SB-93-44, RI-152-93 (hepth/9308154),
E. Alvarez, L. Alvarez-Gaume, J. L. F. Barbon, Y. Lozano,
CERN-TH.6991/93, FTUAM.93/28;
M. Gasperini, R. Ricci and G. Veneziano, CERN-TH.6960/93,
ROM2F/93/24 (hepth/9308112).

\define\KLIMCIK
C. Klimcik and A. Tseytlin, CERN-TH.7069/93 (hepth/9311012).

\define\GERSHON
D. Gershon, TAUP-2121-93, (hepth/9311122).

\define\BERG
E. Bergshoeff, I. Entrop and R. Kallosh, SU-ITP-93-37,
UG-8/93 (hepth/9401025).

\define\AMATI
D. Amati and C. Klimcik, Phys. Lett. {\bf B219} (1989) 443;
G. Horowitz and Steif, Phys. Rev. Lett. {\bf 64} (1990) 260;
Phys. Rev. {\bf D42} (1990) 1950.

\define\TEIT
M. Banados, C. Teitelboim and J. Zanelli, Phys. Rev. Lett. 69 (1992)
1849.

\define\ALI
A. Ali and A. Kumar, Mod. Phys. Lett. A8 (1993) 2045;
G. Horowitz and Welch, Phys. Rev. Lett. 71 (1993) 328.

\define\BIRRELL
N. Birrell and P. Davies, Cambridge University Press (1982).

%%%%%%%%%%%%%%%%%%%%%%%%%%%%%%%%%%%%%%%%%%%%%%%%%%%%%%%%%%%%%%%%
\hfill\vbox{\hbox{IP/BBSR/94-02}
\hbox{IMSC/94-02}\hbox{January, 1994}}\break

\title{\bf{Exact Duality and Nilpotent Gauging}}
\author{Alok Kumar}
\address{Institute of Physics\break
Bhubaneswar-751005, India}
\centerline{and}
\author{Swapna Mahapatra
\foot{e-mail:
kumar@iopb.ernet.in, swapna@imsc.ernet.in}}
\address{Institute of Mathematical Sciences\break
C. I. T. Campus, Madras-600113, India}

\abstract
We obtain new duality transformations, relating some
exact string backgrounds, by defining the nilpotent duality.
We show that the ungauged $SL(2,R)$ WZW model
transforms by its action into the three dimensional
plane wave geometry.
We also give the inverse transformation
from the plane wave to the $SL(2, R)$ model and
discuss the implications of the results.

\endpage
%%%%%%%%%%%%%%%%%%%%%%%%%%%%%%%%%%%%%%%%%%%%%%%%%%%%%%%%%%%%%%%%%%
The study of the classical solutions in string theory
has received much
attention in recent times. There have been two aspects of this study.
The first one is to investigate
the solutions of the low energy classical action \con\GIBB
\noc and the second one to find the exact solutions
\WITT\DVV\HORO\SWAP.
The solutions of the second type are more interesting from the
string point of view
as they correspond to the two dimensional conformal field theories.
The Wess-Zumino-Witten model is one of the examples of such a
solvable conformal field theory.  Many
interesting solutions as exact conformal backgrounds
have been obtained, the first one being the celebrated two dimensional
black hole solution of Witten, which has been obtained as a gauged
$SL(2, R)$ WZW model \WITT. It was shown there that the conventional
gauging (axial or vector) of a $U(1)$ subgroup of the $SL(2, R)$
WZW model leads to a two dimensional singular target space having the
structure of a black hole solution. This has been generalized to
higher dimensions also. For example, ${SL(2, R)\otimes U(1)\over{U(1)}}$
theory corresponds to a three dimensional black string solution\HORO.
Many of these solutions are related by
$O(d, d)$ transformations \con\BUSH\VENE\ASHOKE\KKK\KUMAR
\MAHARANA\HHS\noc.

For
the $SL(2, R)$ theory, it is known that the axial and vector gauged models
are related to each other via duality transformations \KIRITSIS
and are
equivalent as conformal field theories though the target space
geometry is very different in both the cases.
%In the gauged $SL(2, R)$ WZW
%model, if one chooses the subgroup $H$ to be compact, then the coset
%manifold $G/H$ describes an Euclidean two dimensional black
%hole (here the generators of the subgroup $H$ is $\sigma_2$, where
%$\sigma_i$'s are the Pauli matrices. If one chooses the subgroup $H$
%to be noncompact, then one obtains the Lorentzian two dimensional
%black hole solution (here the generators of the subgroup $H$ is
%$\sigma_3$).
The generalizations of these results to larger
duality groups such as $O(d, d, Z)$ \GIVEON have also been done. In
these cases, the dual backgrounds were found corresponding to
the isometry groups which were simply translations of some of
the coordinates.
%Explicit expressions
%for the duality transformations upto leading order in $\alpha'$
%are known and there is evidence that duality transformations receive
%corrections in higher order in $\alpha'$ \PANV.
Recently, the
exact duality transformations have also been found for a number
of cases \KLIMCIK\GERSHON\BERG.  The generalizations to the nonabelian
isometry groups \GAUME have been considered as well.

In this paper, we investigate the duality with respect to the
nilpotent subgroups.
Earlier, it has been pointed out \DVV\ that
one can gauge the parabolic subgroup of $SL(2, R)$
in a consistent manner. This gives rise to the
Liouville theory, where two of the degrees of the freedom are removed
by a constraint and the resulting target space is one dimensional.
This result has been confirmed through an explicit
construction in \ARDALAN, where
a one dimensional target manifold was obtained
by gauging the $SL(2, R)$
WZW model by its nilpotent subgroup $E(1)$.

Motivated by this, we have investigated the relationship between
the duality transformations and the nilpotent gauging and
show that the three dimensional plane wave solution can be
related to the original $SL(2, R)$ WZW model by its action.
It is interesting to note that both the initial and the final
solutions in this case are exact string backgrounds. Duality
transformations relating exact string backgrounds have been
recently presented in another contexts \KLIMCIK \GERSHON
\BERG. For example in
\KLIMCIK, plane wave solutions in arbitrary dimension were
presented as duality invariant class. However, our
duality transformations relate the
solutions within the class of \KLIMCIK to the ones outside.

The way one finds the dual background is the following:
one first gauges the isometry and adds to the action a Lagrange
multiplier term. The gauge fields are nondynamical in the theory in the
sense that they do not have a kinetic term. If one integrates out the
Lagrange multiplier term, one gets back the original model. On the
other hand, by integrating by parts the Lagrange multiplier term
and then integrating out the gauge fields, one obtains the dual
action, in which the Lagrange multiplier is a new dynamical fields
\ROCEK.
Similar techniques have been used for nonabelian duality also, but
there the issues are somewhat more involved \GAUME.

The plan of the paper is  the following:
we shall consider the nilpotent gauging of the
$SL(2, R)$ WZW model, where we gauge the subgroup
$E(1)$. We shall use the standard duality prescription to obtain
the dual backgrounds for both the axial and the vector gauged action.
Interestingly, we find that the dual geometry
corresponding to the $SL(2, R)$ WZW model is a three dimensional
target space, representing a plane wave solution. Such plane wave
solutions have been discussed recently by various people in the
context of WZW models based on non-semi-simple groups \KIRIT.
The dual metric we have obtained has two
isometries and a covariantly constant null killing vector.
We then apply the duality inversion with respect to these two
isometries and obtain the original ungauged WZW model for one of
these.

As will be discussed in the end, our duality transformations do not
belong to any of the class of transformations, approximate or
exact, presented earlier\KLIMCIK \GERSHON \BERG.
The reason being the appropriate use
of the null isometry, as suggested in \KLIMCIK,
for generating new backgrounds.

We now start by
considering  the WZW model based on the group $G = SL(2, R)$.
We parametrize the group manifold by,
$$
g = \pmatrix{a & u \cr -v & b},\qquad
 with \quad ab + uv = 1.\eqn\one$$

The group action for the axial gauging is given by
$g\rightarrow h g h$,
where $g\in G$, $h\in H$ and $H$  is the abelian subgroup of
$G$. On the other hand, the group action for the vector gauging
is given by,
$g\rightarrow h g h^{-1}$.
Here, we shall consider the gauging of the nilpotent subgroup
$E(1)$ of $SL(2, R)$, where the subgroup is generated by
$\sigma^+ = \sigma_3 + i \sigma_2$.
The term nilpotent reflects the
fact that $(\sigma^+)^2 = 0$.
Let us first consider the axial gauged model. The gauged action in
this case can be written as,
$$
S_a(g, A) = S(g) + {k\over 2\pi}\int d^2 z Tr(\bar A g^{-1} \partial g
+ A \bar\partial g g^{-1} + A\bar A + g^{-1} A g \bar A)\eqn\two$$
where, $S(g)$ is the ungauged $SL(2, R)$ WZW model action, which is
given by,
$$
S(g) = {k\over 4\pi}\int_{\Sigma} d^2 z Tr(g^{-1} \partial g g^{-1}
\bar\partial g)- {k\over 12\pi}\int_B Tr(g^{-1} d\,g\wedge g^{-1} d\,g
\wedge g^{-1} d\,g)\eqn\three$$
$B$ is a three manifold, whose boundary is $\Sigma$. The gauge
fields $A$ and $\bar A$ take values in the algebra of $H$ and they
transform as,
$$A\rightarrow h(A + \partial)h^{-1}, \quad
\bar A\rightarrow h^{-1}(\bar A + \bar\partial)h.\eqn\four$$
In terms of the above parametrization, the ungauged WZW model action
becomes,
$$
S(g) = -{k\over 4\pi}\int d^2 z(\partial u \bar\partial v + \bar
\partial u \partial v + \partial a \bar\partial b + \bar\partial
a \partial b) + {k\over 2\pi}\int d^2 z \log u(\partial a \bar\partial b
- \bar\partial a \partial b).\eqn\five$$
If $\epsilon$ is an infinitesimal gauge  transformation parameter,
then the above local symmetry is generated by,
$$
\eqalign{\delta g &= \epsilon \,g + g \, \epsilon; \cr
\delta A_i &= - \p_i \epsilon.}\eqn\six$$
We now gauge,
$$
\eqalign{\d a &= \epsilon (2 a - u - v); \cr
\d b &= \epsilon (-2 b - u - v); \cr
\d u &= \epsilon (a + b); \cr
\d v &= \epsilon (a + b).}\eqn\seven$$
The gauge invariant parameters for this case are
$x = u - v$ and $w = a - b
- u - v$ and the complete gauged action can be written as,
$$
\eqalign{S(g, A) &= S(g) + {k\over 2\pi}\int d^2\,z \lbrack \bar A(b \p a +
u \p v + v \p a - a \p v - v \p u - a \p b - b \p u + u \p b) \cr
&+ A(b \bp a + v \bp u - b \bp v + v \bp b + u \bp a - a \bp u - u \bp
v - a \bp b) \cr
&+ A \bar A(-a + b + u + v)^2\rbrack.}\eqn\eight$$

We fix the gauge by choosing $a + b = 0$. So the gauge fixed
action has the form,
$$
\eqalign{S(g, A) &= -{k\over 4\pi}\int d^2 z (\p u \bp v + \bp u \p v -
2 \p a \bp a)\cr
&+ {k\over 2\pi}\int d^2 z \lbrack -{1\over 2} \bar A(x \p w - w \p x) +
{1\over 2} A
(x \bp w - w \bp x) - w^2 A\bar A\rbrack.}\eqn\nine$$
After integrating out the gauge fields one obtains,
$$
S = {k\over 2\pi}\int d^2 z \lbrack{\p w \bp w\over w^2}
                                \rbrack.\eqn\ten$$
The background
metric, antisymmetric tensor and the dilaton fields
corresponding to the action \ten can be written as
$$
\eqalign{d s^2 &= {k\over 2} {(d w)^2\over w^2}, \cr
B &= 0},\eqn\eleven$$
and
$$
\Phi = - \log w + constant.\eqn\twelve$$
This is just the $SL(2, R)/E(1)$ gauged WZW model\ARDALAN.
The resulting target space is  one dimensional
and the effective action corresponds to the Liouville action.

Now we obtain the model dual to the $SL(2, R)$ WZW model
with respect to its nilpotent subgroup.
For this, we start with the
original $SL(2, R)$ WZW action, gauge it and then add the Lagrange
multiplier
term to the action. The gauged action with the Lagrange multiplier
term (which is gauge invariant) is given by,
$$
S(g, A, \lambda) = S(g) + {k\over 2\pi}\int d^2 z \lbrack\bar A J +
A \bar J -
w^2 A \bar A\rbrack + {k\over 2\pi}\int d^2 z (A \bp\lambda -
\bar A \p\lambda)
\eqn\twelv$$
where the currents $J$ and $\bar J$ are given by,
$$
\eqalign{ J &= \lbrack b \p a + u \p v + v \p a - a \p v -
v \p u - a \p b - b \p u + u \p b\rbrack ;\cr
\bar J &= \lbrack b \bp a + v \bp u - b \bp v + v \bp b + u \bp a
- a \bp u - u \bp v - a \bp b\rbrack}\eqn\thirt$$
and $\lambda$ is the Lagrange multiplier. The gauge
fixed action is obtained by putting the condition
$a + b = 0$. Now fixing the gauge and integrating out the gauge fields,
the dual
sigma model action is obtained in terms of the gauge invariant
parameters and $\lambda$. The expression is given by,
$$
S = {k\over 2\pi}\int d^2 z \lbrack {\p w \bp w\over{w^2}} - {\p \lambda
\bp\lambda\over{w^2}} -
{1\over 2} {\p\lambda\over w^2} (x\bp w - w\bp x) +
{1\over 2} {\bp\lambda\over {w^2}}
(x\p w - w\p x)\rbrack\eqn\fourt$$
where we have used the values of $A$ and $\bar A$ as (obtained by
using their equation of motion),
$$
\eqalign{A &= -{1\over w^2}\lbrack \p \lambda + {1\over 2}
(x \p w - w \p x)\rbrack;\cr
\bar A &= {1\over w^2}\lbrack \p\lambda + {1\over 2}(x \bp w -
w \bp x)).}\eqn\fift$$

{}From the above expression, we can read off the values for the dual
background metric $\t g_{i j}$ and antisymmetric tensor field $\t b_{i
j}$. The dual metric is
given by,
$$
d\tilde s^2 = {k\over{2 w^2}}\lbrack (d w)^2 - (d \lambda)^2
-x d \lambda d w +
w d \lambda d x \rbrack \eqn\sixtn $$
The dual dilaton is found to be,
$$\t\Phi = - \log w + constant,\eqn\sevtn$$ and $\t b_{i j}$ is zero.
Unlike the
${SL(2, R)\over E(1)}$ model, one extra degree of freedom does
not get removed and the target space is three dimensional.
The only non zero
component of the Ricci tensor is $R_{2 2} = - {2\over w^2}$. The
curvature scalar  $R$ is zero as $g^{22}$ is zero (here $w \equiv 1$,
$\lambda \equiv 2$ and $x \equiv 3$).
These backgrounds satisfy the
one loop beta function equations namely,
$$\eqalign{R_{\mu\nu} + 2 \nabla_{\mu} \nabla_{\nu}\t\Phi &= 0;\cr
R + 4/k' - 4 (\nabla\t\Phi)^2 + 4 \nabla^2 \t\Phi &=0.}\eqn\eitn$$
We have also explicitly
verified that they satisfy the two loop beta
function equations.

Recently, in a series of papers\BARS,
the path integral formulations of the gauged WZW model
has been given to
compute the exact string backgrounds. It was found that the
quantum action relevant for computing the exact backgrounds
differs from the classical one by a term which
is proportional to $Tr[A\bar A]$. In our case this term is zero due to
the nilpotency of the gauged subgroup. As a result we
expect the backgrounds in eqns. \sixtn-\sevtn to be exact
to all orders in sigma model.

Also, by making a coordinate redefinition,
$$y = {x\over w}\eqn\nintn$$ which is well defined, since
$w \neq 0$, we can rewrite the dual metric as,
$$
d\t s^2 = {k\over 2} \lbrack {1\over w^2}((d w)^2 - (d \lambda)^2) +
d \lambda d y\rbrack\eqn\twenty.$$
The metric $\t g_{i j}$ now is independent of $\lambda$
and $y$.
After shifting $y\rightarrow 2 y$, we obtain,
$$
d\t s^2 = {k\over 2}\lbrack{1\over w^2}((d w)^2 - (d \lambda)^2) +
2 d \lambda d y\rbrack\eqn\twentyo$$
which corresponds to a plane wave
solution.
The killing vector (in $\lambda$-direction)
$k_a = (0, 1, 0)$ is a null vector as
$k_a k^a = 0$ and $k_{a; n} = k_{a, n} -
\Gamma^{\lambda}_{a n} k_{\lambda}
= 0$. This implies that $k_a$ is
a covariantly constant null killing vector. The existence of such
a vector and the form of the metric implies that the dual background
is basically a three dimensional plane wave solution, which is
known to be an exact solution of string theory to all orders \AMATI.
This is consistent with the arguments presented above.

We have also analyzed the vector gauged dual model.
The vector gauged action is given by,
$$
S_v(g, A) = S(g) + {k\over 2\pi}\int d^2 z Tr(A \bp g g^{-1} -
\bar A g^{-1} \bp g + A \bar A - g^{-1} A g \bar A).\eqn\twentt$$

The vector gauge transformation is given by, $g\rightarrow h g
h^{-1}$. The local vector symmetry is generated by,
$\d g = g \,\epsilon - \epsilon \,g$. We gauge the symmetry,
$$
\eqalign{\d a &= \epsilon (v - u);\cr
\d b &= \epsilon (u - v); \cr
\d u &= \epsilon (a - b - 2 u);\cr
\d v &= \epsilon (-a + b + 2 v).}\eqn\twenthr$$
Here the gauge invariant parameters are
$x' = a + b$ and $w = a - b - u - v$. We choose the gauge
$u - v = 0$ and again add the Lagrange multiplier term to the
gauged action. The gauge fixed action is given by,
$$
\eqalign{S(g, A, \lambda) &= -{k\over 4\pi}\int d^2 z (2 \p u \bp u +
\p a \bp b
+ \bp a \p b) \cr
&+ {k\over 2\pi}\int d^2\,z\lbrack -{1\over 2} A (x' \bp w - w
\bp x')
- {1\over 2} \bar A(x'\p w - w \p x') + w^2 A \bar A\rbrack \cr
&+
{k\over 2\pi}\int d^2 z (A \bp\lambda - \bar A \p\lambda)}\eqn\tewntfo$$
Integrating out the gauge fields as before we obtain,
$$
S = {k\over 2\pi}\int d^2 z\lbrack {\bp w \p w\over w^2} + {\bp \lambda
\p \lambda\over w^2} + {x'\over 2 w^2} (\p \lambda\bp w + \bp\lambda
\p w) - {1\over 2 w} (\p\lambda \bp x' + \bp \lambda \p x')\rbrack.
\eqn\twentfiv$$

{}From the above expression, we find that the dual antisymmetric tensor
field is again zero and the dual metric is given by,
$$d\t s^2 = {k\over 2}\lbrack{1\over w^2}((d w)^2 + (d \lambda)^2) +
{x'\over  w^2} d \lambda d\, w - {1\over  w} d \lambda d\,x'
\rbrack.\eqn\twentsix$$
The dual dilaton is again given by,
$$
\t \Phi = -\log w + constant.\eqn\twentsev$$
As in the case of axial gauging case, we define, $y = {x'\over w}$
and shifting $y\rightarrow 2 y$ we obtain,
$$
d\t s^2 = {k\over 2}\lbrack {1\over w^2} ((d w)^2 + (d \lambda)^2) -
2 d \lambda d y\rbrack.\eqn\twenteit$$

This is again a plane wave solution in three dimensions.
We note that the metric in the two different gauged models,
eqns.\twentyo and \twenteit, are esentially same upto analytic
continuations.  This follows from the fact that $A\bar A$ term
is zero in eqns.(2) and \twentt. As a result, the gauged actions
are related by a field redefinition $\bar A \rightarrow - \bar A$.
Taking this into account the dual metric in two
different gauging, after a shift $\lambda \rightarrow \lambda -
y$, become (upto analytic continuation),
$$d\t s^2 = {k\over 2}\lbrack{1\over w^2}((d w)^2 - (d \lambda)^2) -
(d y)^2 ({1 + 2 w^2\over w^2}) + 2 d \lambda d y ({1 + w^2\over w^2})
\rbrack\eqn\twentnin$$

Now, if we apply the duality inversion \BUSH\ROCEK with
respect to the
isometry in $y$ direction for the metric \twentnin, we get,
$$\eqalign{d\t s'^2 &= {k\over 2}\lbrack{1\over w^2}(d w)^2 -
({w^2\over{1 + 2 w^2}})
(d y)^2 + ({w^2\over{1 + 2 w^2}})(d \lambda)^2\rbrack;\cr
\t B' &= -1;\cr
\t \Phi' &= 2 \phi + constant.}\eqn\thirtfiv$$.

More interesting is the case of
$\lambda$ isometry for the metric \twentnin for which
$$\eqalign{d\t s'^2 &= {k\over 2} \lbrack {1\over w^2}(d w)^2 - w^2 (d
\lambda )^2 +
w^2 (d y)^2 \rbrack;\cr
\t B'_{\lambda y} &= -(1 + {w}^2);\cr
\t {\Phi'} &= constant.}\eqn\thirtfour$$

Interestingly
the final background in \thirtfour is  same as the ungauged
$SL(2, R)$ WZW model. To show this we first find, by an explicit
computation, that the Ricci tensors $R_{\mu\nu}$ for the metric
in eqns.\thirtfour is proportional to $g_{\mu\nu}$. As a result
this is the metric for a maximal symmetric space.
More explicitly,
by coordinate reparametrizations and analytic continuations,
the metric in eqn.\thirtfour can be
transformed to the Robertson-Walker form in three dimensions:
$$
d\t s'^2 = {k\over 2}\lbrack  (d t)^2 + exp(2t)
[(dr)^2 + r^2 (d \theta)^2] \rbrack\eqn\thirtysev;$$
By another coordinate transformation, similar to the one in four
dimensional case\BIRRELL,
this can be written in the static form:
$$
d\t s'^2 = {k\over 2}\lbrack {(1-r^2)} (d t)^2 +
 {(1-r^2)}^{-1}(dr)^2 + r^2 (d \theta)^2 \rbrack\eqn\thirtyeit.$$
which, after some analytic continuations, can be identified with
the metric of the three dimensional black hole \TEIT\ALI.
Together with the antisymmetric tensor and the dilaton as
in eqn.\thirtfour, it
can be interpreted as the ungauged SL(2, R) WZW model \ALI.

To conclude, in this paper we have presented the duality
transformations which
relate two exact solutions of string theory,
namely the WZW models and the plane wave sulutions.
We would like to point out once again that
the duality transformations applied in our case are outside the set
of $O(2, 2)$ transformations \KUMAR\GERSHON that are normally
considered for a one coordinate dependent background in three
space-time dimensions. For generating inequivalent backgrounds,
the $O(2, 2)$ transformations have only one nontrivial
parameter \ASHOKE. One can use this to transform away
the antisymmetric tensor present in the original ungauged WZW
model. However, it turns out that this demand fixes the
background for other fields as well. The final model in this
case is a product space of the 2-d black hole
with a flat coordinate \KUMAR\GERSHON. In
the present case we once again have a vanishing antisymmetric
tensor. But the other backgrounds are different than the  product
space just discussed.
Hence the plane wave solution does not belong
to the class of models related to the
ungauged SL(2, R) model by $O(2, 2)$ transformations. A more likely
interpretation of our transformations is probably in terms of a
suggestion presented in \KLIMCIK where it was pointed put that
the duality transformation, after mixing the null isometry
with the other ones, can give rise to new consistent backgrounds.
It will be interesting to examine this proposal further. It
will also be interesting to examine
these results for the higher dimensional plane wave solutions
and WZW models as well.

\noindent{\bf{Acknowledgement}:} One of us (S. M.) would like to
thank Institute of Physics,\break Bhubaneswar for its hospitality, where
part of this work was done.

\refout
\end